# Enhancing Inventory Management with Progressive Web Applications (PWAs): A Scalable Solution for Small and Large Enterprises

Abhi Desai
New England College, Henniker, New Hampshire, US
*adesai_gps@nec.edu | desai.abhi94@gmail.com*

*Abstract*—Efficient inventory management is crucial for both small and large enterprises to optimize operational workflows and reduce overhead costs. This paper explores the development and implementation of a Progressive Web Application (PWA) designed to enhance the inventory management experience. The application integrates key functionalities such as barcode and QR code scanning, geolocation-based warehouse identification, and cross-device accessibility. By leveraging PWA technology, the solution ensures offline capabilities, responsive user experience, and seamless adaptability across various platforms. The study discusses the challenges and benefits of implementing PWA in inventory management systems, including its limitations in performance compared to native applications. Insights from the development process provide a roadmap for future developers looking to integrate PWA technology into enterprise applications. This research contributes to the growing domain of web-based inventory solutions, offering a scalable and cost-effective alternative to traditional inventory management software.

I. INTRODUCTION

*A. Problem Characteristics*

Warehouse resource management is a key element of the effective functioning of many enterprises, especially in industries related to logistics, trade, and production. This problem becomes particularly important in the context of growing expectations related to process automation and the minimization of errors resulting from manual management.

Inventory management is crucial for the efficient operation of many businesses, particularly in the logistics, trade, and manufacturing sectors. This issue gains prominence as companies face increasing demands for process automation and a reduction in errors due to manual management. The challenge is exacerbated for new market entrants that often lack the necessary capital to develop specialized software for every device type.

*B. Goal Header*

Our goal is to create a tool that will help companies track and manage assets more efficiently. The main technical assumption was to create a PWA application in the MVP (Minimal Viable Product) version, i.e. with CRUD functions, enabling the management of warehouses, stock levels, product availability and assigned product categories. The functionality of the application has been extended to include a QR code and barcode scanning module. Another element is the implementation of geolocation, which aims to accelerate the identification of the warehouse based on the user's location. Our goal is to create a tool that will help companies track and manage resources more effectively. The main technical assumption was to create an MVP (Minimal Viable Product) version of the PWA application, i.e., one that has CRUD functionalities to manage warehouses, inventory, product availability, and assigned product categories. The goal of creating a PWA application was to create a single tool that could be used on many devices, regardless of whether the device is running on the preferred system. The application's functionality was supposed to be extended with a QR code and barcode scanning module. Scanning the codes was to enable faster obtaining of information about a given product. Another sensor-based functionality is geolocation, which can speed up warehouse identification based on the user's location. QR code scanning was to be extended to edit the warehouse status after scanning the QR code. The QR code was to contain data that the system would read and present to the employee for processing. This solution could significantly speed up the receipt and issuance of products from warehouses.

*C. The Benefits*

In the business context, this project sought to enhance resource management efficiency, decrease the time required for warehouse operations, and minimize errors stemming from manual tasks. On the technical side, the main objective was to develop a user-friendly and effective tool that seamlessly integrates with current enterprise systems and is and is usable on mobile devices and computers. In a business setting, the project sought to develop a user-friendly and effective tool that is readily accessible on both mobile devices and computers. Additionally, the system aimed to reduce mistakes associated with manual inventory management.

## II. THE RELATED WORKS

Enterprise Resource Planning (ERP) systems, such as SAP [1], Oracle NetSuite [2], and Odoo [3], are popular solutions in this domain offering comprehensive features ranging from inventory tracking to financial management. However, these platforms often require significant resources for deployment and maintenance, making them less accessible for smaller enterprises.

For small- and medium-scale enterprises, SaaS (Software as a Service) solutions like Zoho Inventory [4] and TradeGecko [5] are popular alternatives. These tools are user-friendly, lightweight, and integrate well with other software, but they lack scalability and deep customization for larger operations. The above-mentioned software is mainly used to process orders and ship items from given warehouses. We want to investigate whether achieving similar functionality, as well as automating some operations (receiving and issuing products) is possible using PWA (Progressive Web Application). PWA technology offers several significant advantages that can translate into greater flexibility and lower maintenance costs compared to classic SaaS solutions:

- Lower maintenance costs - PWAs do not require maintaining separate versions for different platforms (e.g., iOS, Android, desktop).
- Offline operation and resource savings - PWA can operate offline thanks to Service Worker's technology, which means that basic functions are available even with limited internet access.
- Independence from the App Store and Play Store - No need to publish applications in app stores eliminates the associated costs and waiting time for acceptance.

In the context of technology, barcode and QR code integration has become a cornerstone of modern inventory systems. Libraries such as ZXing [6] offer reliable and efficient methods for decoding these codes, making product identification and tracking much easier.

Progressive Web Apps (PWAs) [7] are gaining traction for their ability to combine the accessibility of web applications with the native performance of mobile apps. Their features include offline functionality, simple installation, and access to device capabilities such as cameras and geolocation. PWAs also enable unified development across platforms, reducing costs and maintenance efforts.

## III. THE RESULTS

*A. The Summary*

The project implementation was successful. We included all important functionalities and comprehensively checked the usability and ease of use of PWA technology.

Our project builds upon these insights to deliver a PWA-based inventory management system tailored for enterprises of varying sizes. By integrating technologies such as Angular [8], Spring Boot [9], PostgreSQL [10], and AWS [11], we aim to create a scalable, cost-effective, and accessible solution that combines the simplicity of SaaS tools with the power of ERP platforms.

The implementation of the project proceeded as planned and was successful. We managed to include all important functionalities and use of PWA technology.

*B. The Project Assumptions*

Our use of PWA implements on multiple devices while maintaining a single code base. At the same time, it provides access to device sensors, like the camera or geolocation, which we used to add additional functionalities, such as scanning bar codes and QR codes and pinpointing the exact warehouse the user is currently stationed in. This makes it especially helpful for businesses with multiple warehouses.

## IV. THE FUNCTIONALITIES

- CRUD operations, i.e., Create, Read, Update, Delete, allowing full modification of data stored in the database.,
- Ability to scan barcodes and QR codes in an intuitive way for quick access to a specific product in the warehouse,
- Inventory management in multiple warehouses,
- Managing users and their access to functionalities offered by the system layer
- Create, Read, Update, and Delete (CRUD) operations, allowing for full modification of data stored in the database, such as items, warehouses, inventor,y and item categories,
- Ability to scan barcodes and QR codes in an intuitive way to quickly access a specific product in stock,
- Inventory management across multiple warehouses,
- Management of users and their access to functionalities offered by the system layer,

*A. The Business Goals*

1) Enhanced Inventory Management: The system integrates barcode and QR code scanning to improve asset tracking, streamline inventory control, and provide quick access to details such as stock levels and locations.

2) Improved Efficiency: Simplified management of incoming and outgoing deliveries enhances organization, reduces processing times, and boosts overall productivity.

*B. The Technical Goals*

1) Cross-Platform Compatibility: Using Progressive Web App (PWA) technology ensures the app works seamlessly across devices while maintaining a single codebase, reducing development and maintenance efforts.
2) Scalable Architecture: The backend, designed with Spring Boot and deployed on AWS using Terraform, is built to handle increasing data loads and user demands.
3) Secure Data Management: The app protects user data and ensures resource access authorization by using Amazon Cognito for authentication and other security measures.
4) Modern Development Practices: Technologies like Terraform, Docker, and Nginx highlight the use of efficient, modern tools for development, deployment, and infrastructure management.
5) Testability: The software that was developed is comprehensively tested with over 80% code coverage.

*C. The TechStack*

Our project uses a modern and robust tech stack designed to deliver both scalability and reliability: Angular [8], Spring Framework [9], AWS (Amazon Web Services) [11], Docker [12], Hashicorp Terraform [13].

Moreover, we used: ZXing: A library integrated into our application to facilitate barcode and QR code scanning. Utilizing the device's camera, we implemented ZXing to efficiently read both QR codes and EAN-13 barcodes, ensuring accurate and seamless data capture directly within the application. This approach leverages PWA capabilities to access hardware resources, eliminating the need for external scanning devices and enhancing usability on various platforms. [6]

*D. The Project Assumptions*

At the beginning of the project, we established various constraints and assumptions to limit the scope and guide the development in the right direction.

Our goal was to create a simple and intuitive application that would allow users of various skill levels to successfully use our application. Thanks to PWA technology, we wanted to obtain an application available on many devices, with simple installation and a native graphical interface.

We decided to limit the scope and support only one company with multiple warehouses. These warehouses may contain items shared between them and have individual inventories. There will be no registration form - each employee in the company will have to be added manually by the administrator

At the beginning of the project, we set multiple restrictions and assumptions to limit the scope and guide the development in the right direction.

We set out to create a simple and intuitive application to allow users across all skill levels to successfully use our application. Thanks to the PWA technology,y we wanted to achieve an application that is available on multiple devices with simple installation and a native feel.

We decided to limit the scope and support only a single company owning multiple warehouses. Those warehouses can contain items shared between them and have individual stocks. There will not be a registration form—the administrator will have to manually add each employee to the company. This will ensure that only users declared by the companies will use the application, reducing the cost of maintaining it.

*E. The Additional Highlights*

- Feasibility Demonstration: The app proves the potential of PWAs for small and medium-sized companies thanks to the quick development process. Additionally, the use of hardware sensors makes it a good solution for an inventory management application, allowing for quality-of-life features like scanning of the QR codes.
- Technical Design: Built with scalability and security in mind, the system employs modern tools such as AWS and Terraform, aligning with enterprise-level best practices and conventions. It makes use of AWS infrastructure, and through the usage of private subnets, appropriately applied security groups, and Amazon Cognito, the application is secured, and thanks to the application load balancing and auto-scaling, the application is scalable without generating unnecessary costs.

## V. THE CONCLUSION

To sum up, the application we designed is intuitive to use and provides basic functionalities required by warehouse workers. It will allow for more efficient and simpler inventory management. Thanks to PWA technology, it is available on both mobile devices and computers, supporting installation and quick and easy access in conditions of limited Internet access.

What is important for the technological recipient is that we have tested the capabilities of PWA technology and proved that it is profitable for creating applications for multiple devices while maintaining one code base.

In summary, the app we developed uses PWA technology to enable more flexible and cost-effective inventory management. Unlike traditional solutions that require dedicated barcode scanning devices, this app can be easily used on standard smartphones, making it a more attractive option for small businesses.

With PWA technology, the app works seamlessly on both mobile and desktop devices, eliminating the need for separate apps for different platforms. It runs as a native app, providing access to device resources such as camera and location services, and can maintain functionality even with limited or no internet access. This cross-platform compatibility, combined with reduced hardware requirements, makes it a practical solution for modern inventory management.

## VI. THE FURTHER RESEARCH

For further development of the project, it would be necessary to take into account what other functionalities may be useful for employees and warehouse managers, e.g.

- optimization of the arrangement of items in the warehouse,
- using geolocation to determine the shortest route to the desired section in the warehouse,
- Calculating route and cart capacity requirements to complete an order from the warehouse,
- expanding the application to a commercial version that can support many companies in a comprehensive way,
- and many others.

For further development of the project, it would be important to consider what other functionality might be useful for warehouse workers and managers, such as:

- optimization of the layout of items in the warehouse,
- use of geolocation to determine the shortest route to the desired section in the warehouse,
- calculation of the route and cart capacity requirements to complete an order from the warehouse,
- expansion of the application to a commercial version that can support multiple companies in a comprehensive way
- handling deliveries and releases of products from the warehouse
- managing accounts through the company
- supporting inventory processes
- keeping statistics and generating appropriate reports and charts based on the data
- enabling integration with other systems to automate the processes of deliveries and releases from warehouses (adding generation of QR codes and invoices) or generating QR codes based on invoices.